\documentclass[english]{llncs}
\usepackage[T1]{fontenc}
\usepackage{listings}
\usepackage[latin9]{inputenc}
\usepackage{graphicx}
\usepackage{pgfplots}
\usepackage{url}
\usepackage{babel}
\pgfplotsset{width=6cm,compat=1.18}
\begin{document}
\title{SimPLoID \\ Harnessing probabilistic logic programming for infectious disease epidemiology}
 \author{Felix Weitk\"amper\inst{1} \and Ameen Almiftah\inst{1} \and Kailin Sun\inst{2}}
 \institute{Institut f\"ur Informatik der LMU M\"unchen, Oettingenstr. 67, 80538 M\"unchen,
 Germany \and Department Biologie I der LMU M\"unchen, Menzinger Str. 67, 80638 M\"unchen,
 Germany}
\maketitle
\begin{abstract}
High quality epidemiological modelling is essential in order to combat the spread of infectious diseases. In this contribution, we present SimPLoID, an epidemiological modelling framework based on the probabilistic logic programming language ProbLog. SimPLoiI combines concepts from compartmental modelling, such as the classic Susceptible-Infected-Recovered (SIR) model, with network-based modelling. 
As a proof of concept, SimPLoID showcases the potential of declarative probabilistic logic programming for a natural,  flexible and compact expression of infectious disease dynamics. In particular, its modularity makes it easily extendable in the face of changing requirements.  
This application area benefits especially from the precisely specified semantics of the ProbLog language and from the well-maintained engines, which support a variety of query types from exact inference to Monte Carlo simulation. 
We also provide a domain-specific language designed for researchers not trained in programming, which is compiled to ProbLog clauses within an interactive Python application. 
\end{abstract}

\section{Introduction}

Infectious disease epidemiology is the study of the spread of infectious
diseases. This discipline is highly relevant for public health management,
as important policy decisions (for instance vaccination programmes,
quarantine, and resource distribution) are based on epidemiological
research. Therefore, the quality of data and modelling is essential
for implementing effective decision-making on a population level.
In this paper, we introduce a tool which allows epidemiology researchers
to integrate their assumptions quickly and efficiently into a network-based
model which can be easily adapted to all infectious diseases. 

\begin{figure}
\centering
\includegraphics[width=0.7\linewidth]{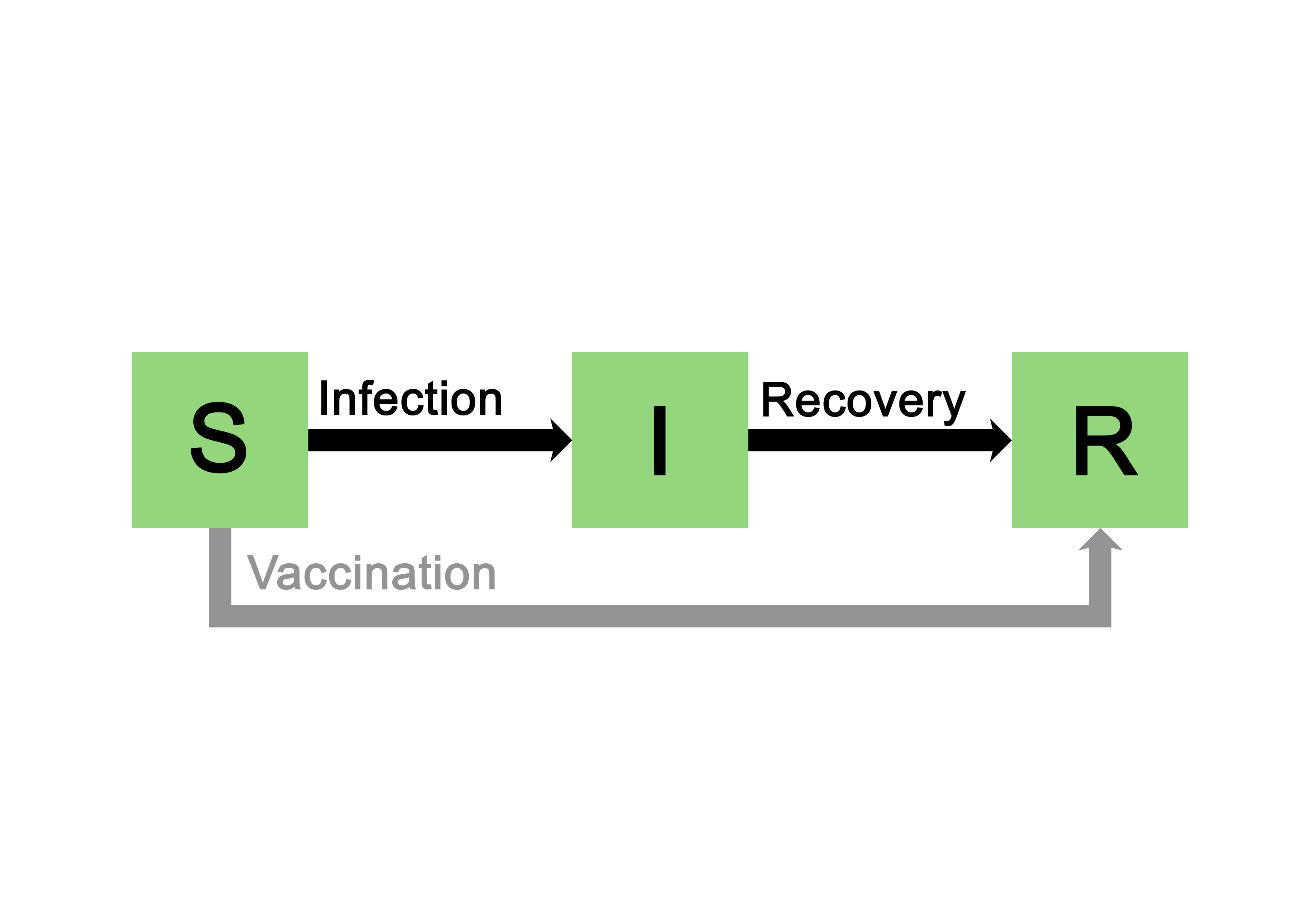}
\caption{The classic SIR model, showing Susceptible, Infected and Recovered
groups with their modes of transition \cite{DBLP:conf/iclp/WeitkamperSS21}. }\label{FigSIR}
\end{figure}

A classical approach to epidemiological modelling is via compartmental models
such as Kermack and McKendrik\textquoteright s Susceptible-Infected-Recovered
(SIR) model, which has been used successfully to model many human, animal and plant diseases \cite{Anderson1991}. This model divides the population into the
three different stages of disease infection \cite{Kermack1927}, summarised
diagrammatically in Figure \ref{FigSIR}. The proportion of the population moving
from being susceptible to infected to recovered in each time step
is modelled by differential equations. 

The classic SIR model relies on some assumptions which mean it may
not be the best model for all infectious diseases. 
For instance, a key assumption is a well-mixed population, 
the violation of which during the SARS-CoV-2 pandemic led to a significant deviation from the theoretical distribution of the illness \cite{Wong2020}.

\begin{figure}
\centering  
\includegraphics[width=1\linewidth]{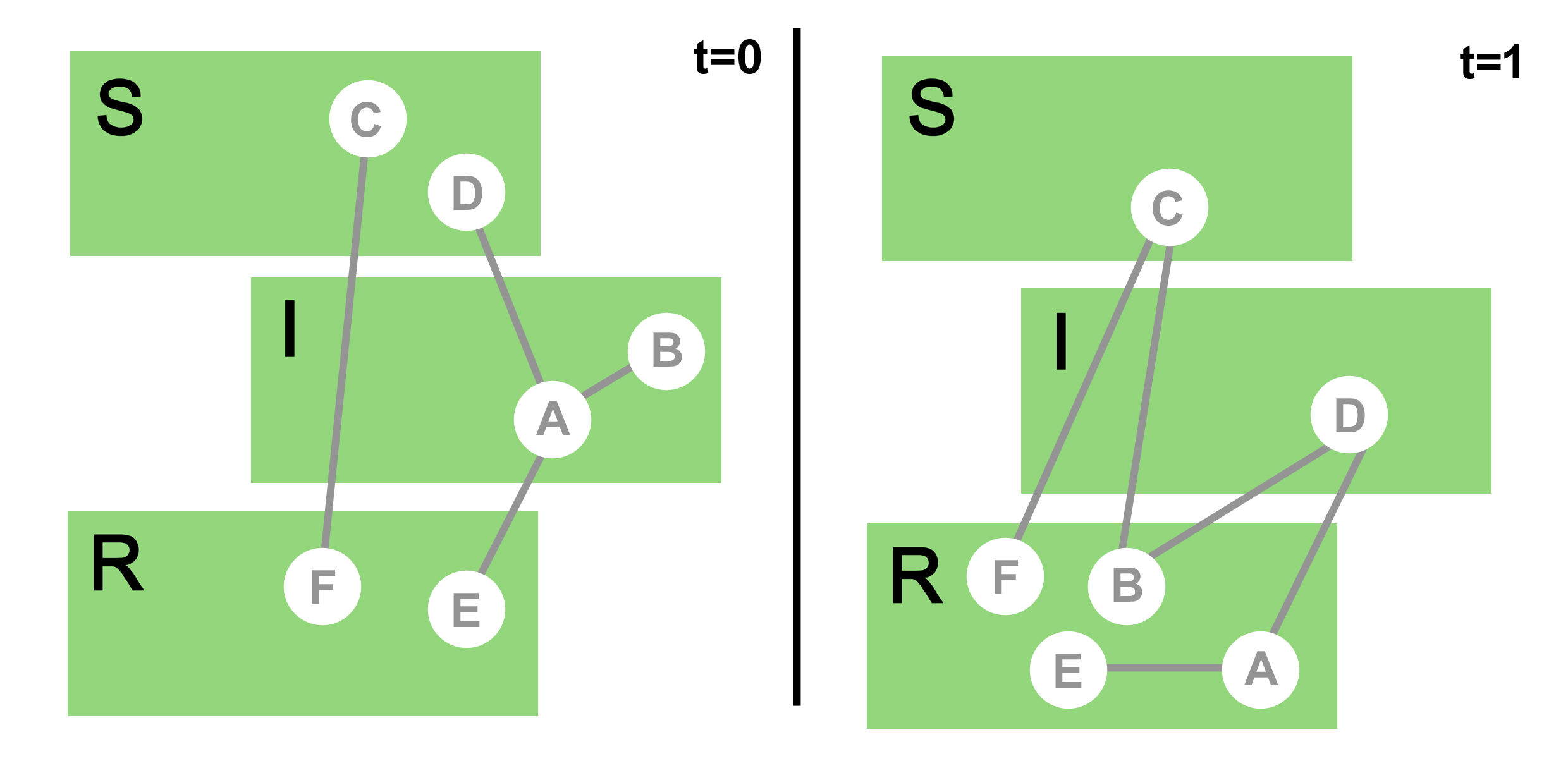}
\caption{An example of a network-based model. 
Circles represent individuals, which may be either susceptible, infected or recovered. Lines represent connections between individuals. 
Across different timesteps, both the compartments of the individuals and the connections between them can shift.}\label{fig:network}
\end{figure}

These limitations can be overcome by a network-based model,
which uses the concept of the SIR model but applies it to a contact
graph where nodes represent individuals and edges show the connection
between individuals, which individually are in one of the compartments of the classical model.  
As connections between individuals can change over time, and through hospitalisation, quarantine or other measures even in direct response to illness, it is important for the underlying connection graph to allow for changes over time, as illustrated by Figure \ref{fig:network}. 

It is important for an epidemiological modelling platform to be flexible, transparent
and versatile in the face of constantly updating information and fast-changing
parameters, for instance during the course of a new epidemic. 
We believe that as a declarative probabilistic modelling paradigm based on clearly stated human-readable rules,
probabilistic logic programming is highly suitable for this purpose.
More specifically, the particularly simple yet expressive 
probabilistic logic programming language ProbLog is
ideal for these purposes, because the parameters and underlying assumptions
of the model are transparent and immediately accessible \cite{deRaedt2007}. 
ProbLog is implemented by the well-maintained ProbLog 2 engine \cite{Fierens2015}, 
which allows not only for the calculation of exact probabilities, but also for simulations to be generated from a specified model \cite{Dries2015};
these useful features allow for the development of a versatile and extendable epidemiological tool.

In this contribution we introduce SimPLoID, short for Simulations with Probabilistic Logic for Infectious Diseases,  a flexible and transparent framework for epidemiological modelling that is also easy to use for researchers who are not trained in programming. However, the current implementation is to be regarded as a proof of concept, showcasing the potential of probabilistic logic programming for disease epidemiology, rather than as production software. An indication of what could be added in future work is given in  Section \ref{sec:Discussion}.  

\section{State of the art}

Our contribution is a modelling framework that is based on probabilistic logic programming.
It serves as a proof of concept for an idea first presented by Weitk\"amper et al. \cite{DBLP:conf/iclp/WeitkamperSS21}.
Of course, given the importance of epidemiological modelling, a variety of tools and packages are available for this purpose. Most of them are written in imperative languages, and do not allow transparent access to modelling parameters beyond the user's original settings. They also lack a clear underlying semantic framework.  

An exception is EMULSION \cite{picault2019emulsion}, which supports both compartmental and network-based approaches. Features of EMULSION include a graphical user interface for simulations as well as statistical tools for analysing simulations. 
EMULSION compiles the model file, written in its purpose-built domain-specific language, into a set of finite state machines, one for every modelled quantity. 
In a network-based model, this means a single finite state machine for every individual in the simulation.
As this concept is closest to our approach, we discuss the differences between probabilistic logic programs and finite state machines as compilation targets in Section \ref{sec:Discussion}.

\section{System description}
Models and queries are processed in two steps. 
First, a model file, written in a dedicated domain-specific language (DSL), is parsed and compiled into ProbLog clauses.
Model files contain simple statements about the disease under investigation.

A model file can be supplemented by default settings, which are contained in a default file, to avoid the strain of repetitive specifications.
Individuals and their network of contacts can either be generated randomly, triggered by a corresponding statement in the model file, or they are specified as an additional file and loaded directly into ProbLog 2.

ProbLog allows for succinct sets of probabilistic rules which match natural human understanding quite closely. 
This is especially true when making use of the various language extensions supported by the ProbLog 2 engine, in particular the support for inhibition effects \cite{Meert2014}.

\begin{lstlisting}[caption = ProbLog code for a flu model with SIR compartments, label={lst:ProbLog}, float, basicstyle=\footnotesize, numbers=left]
:- use_module(library(db)).
:- csv_load('individualsList.csv','person').
:- csv_load('contactList.csv','airborne_contact').

time(N) :-
    between(1,12,N).
flu__susceptible(X,N) :-
    time(N), person(X).
\+flu__susceptible(X,N):-
    time(N), flu__infected(X,N).
0.1::flu__infected(X,M):-
    time(M), N is M-1, \+flu__infected(X,N), 
    flu__susceptible(X,N).
0.8::flu_infected(X,M) :-
    time(M), N is M-1, airborne_contact(X,Y,N), 
    flu_susceptible(X,N), flu_infected(Y,N).
flu__infected(X,M) :-
    time(M), N is M-1, flu__infected(X,N).
\+flu__infected(X,M) :-
    time(M), N is M-7, flu__infected(X,N).
\+flu__susceptible(X,N) :-
    time(N), flu__resistant(X,N).
flu__recovered(X,M) :-
    time(M), N is M-1, flu__infected(X,N), 
    \+flu__infected(X,M).
0.9::flu__resistant(X,N) :-
    time(N), flu__recovered(X,N).
flu__resistant(X,M) :-
    time(M), N is M-1, flu__resistant(X,N).

query(flu__susceptible(X,N)).
query(flu__infected(X,N)).
query(flu__recovered(X,N)).
query(flu__resistant(X,N)).
\end{lstlisting}

An example demonstrating a simple network-based model SIR model is given by Listing \ref{lst:ProbLog}.
We see here that we can load persons and airborne contacts from csv files, rather than crowding our program by having to assert them explicitly. 
The relevant time points are provided using the Prolog-builtin \lstinline{between/3}, which is also available in ProbLog 2, as is the basic Prolog arithmetic used in the remaining clauses.
The clause at Line 7 asserts that being susceptible is the default value, which holds for all persons and time points unless there is a reason for it not to hold. 
Lines 9 and 21 then give the exceptions to this rule: Whenever an individual is either resistant or currently infected, they are not susceptible. 
Line 11 models the extrinsic rate of infection, where there is a 0.1 chance at any timestep for an individual to be infected from a cause outside of the system we are modelling. 
The clause at Line 14 models the intrinsic infection rate caused by airborne contact between a susceptible and an infected individual.
Line 17 models that in principle, if someone is ill in one timestep they will still be ill in the next timestep.
Line 19 then models recovery after a fixed time of infection, the inhibitor with the negated head again "overruling" the previous clause. 
Note that a stochastic duration of illness could also be modelled by replacing these two clauses with a single probabilistic clause. 
Line 23 explains that an individual is recovered if they had been ill in the previous timestep but are no longer ill in the current timestep.  
Line 26 asserts that there is a 0.9 chance of obtaining immunity upon recovery, and Line 28 ensures that immunity is permanent.

Observe in particular the modularity of the ProbLog code: 
Every possible cause for infection can be added and removed separately, without any alteration to the remainder of the program. 
Those causes are then treated as independent possible triggers for the event specified by the head. 
For instance, if an individual has contact with two infected individuals in a given timestep, the program above will calculate three separate and independent grounds for infection, the two contacts and the baseline rate, resulting in an overall infection probability of 
\[
1 - (1 - 0.1)(1 - 0.8)(1 - 0.8) = 0.964
\]
according to the laws of probability. 

The query goals in the final four lines instruct ProbLog 2 to simulate susceptible, infected, recovered and resistant individuals. 
Note that had less goals been queried, the simulation engine would still have had to simulate almost all of the other ground atoms since they are all mutually connected through the program clauses. 
Thus there is little efficiency gain expected from querying only for, say, the infected individuals.  

These clauses are then passed to the ProbLog 2 system, whose Monte Carlo engine \cite{Dries2015} then generates one or more simulation runs.
Their results in turn can be processed by various Python utilities for graphing and tabulating the output.
In our application, we use the mathplotlib package \cite{hunter2007} for graphing.

To further enhance usability, the different functions of our application can be accessed from an interactive shell, which also provides help and documentation on the different available options. 



Several well-known disease models have been implemented using the system. 
While the DSL provides for a large variety of different compartments, including different transmission forms, vaccination regimes and maternally-derived immunity, we illustrate a simple SIR model here such as the one expressed by Listing \ref{lst:ProbLog}.
Its model file will contain a block for each supported compartment, as in the excerpt for "infected" shown in Listing \ref{lst:infected}. 

\begin{lstlisting}[caption = {Fragment of an infection model},label={lst:infected}, float, basicstyle=\footnotesize, numbers=left]
infected transmission 0.8
infected external 0.1
infected period 7
\end{lstlisting}
Here, "transmission" designates the likelihood of an infection at contact, "external" the likelihood for an infection from outside and "period" the duration of an infection. 

It is also possible to include further information such as the number of simulation runs or whether individuals and their contacts should be generated automatically or loaded from a file, and which compartments should be queried. 
However, these metaparameters can also be specified in the interactive shell, which has been found preferable in most cases. 

After running the simulation, the results of the individual runs can be graphed or tabulated, either collectively or individually. 
Figure \ref{fig:graphs} shows different ways of displaying the simulation outcomes, with different periods of immunity:
\begin{figure}
\centering
\begin{tikzpicture}
\begin{axis}[
enlargelimits=false,
xlabel = time (weeks),
ylabel = infected (individuals)
]
\addplot graphics [
xmin=0,xmax=120,
ymin=0,ymax=50,
] {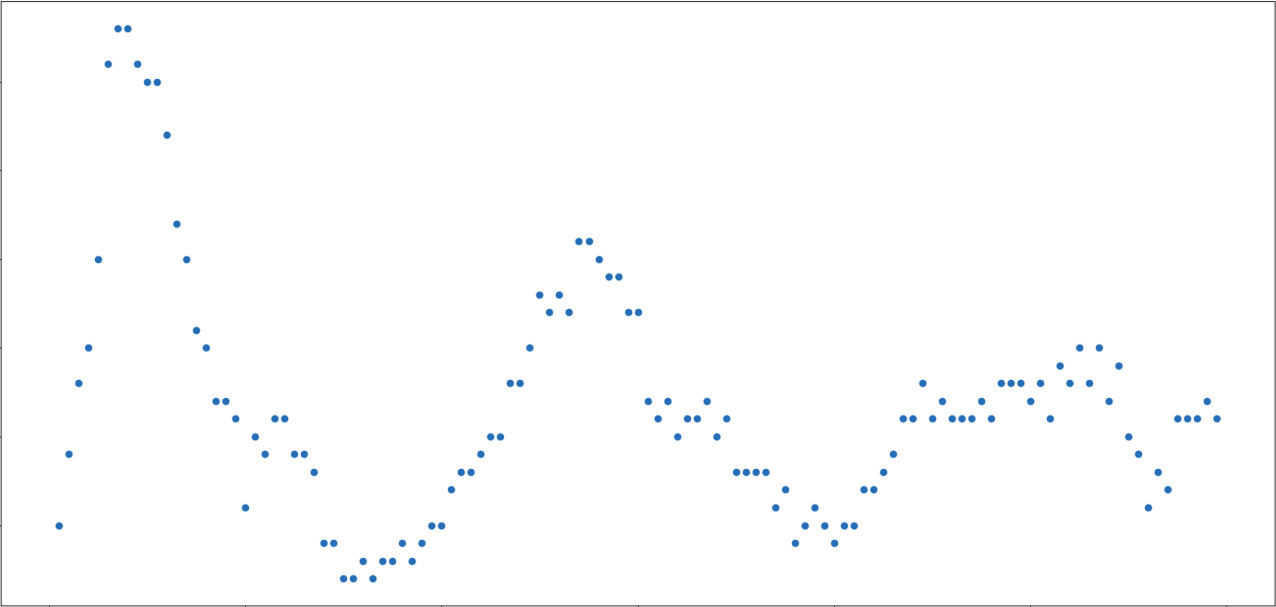};
\end{axis}
\end{tikzpicture}
\begin{tikzpicture}
\begin{axis}[
enlargelimits=false,
xlabel = time (weeks),
ylabel = infected (individuals)
]
\addplot graphics [
xmin=0,xmax=120,
ymin=0,ymax=50,
] {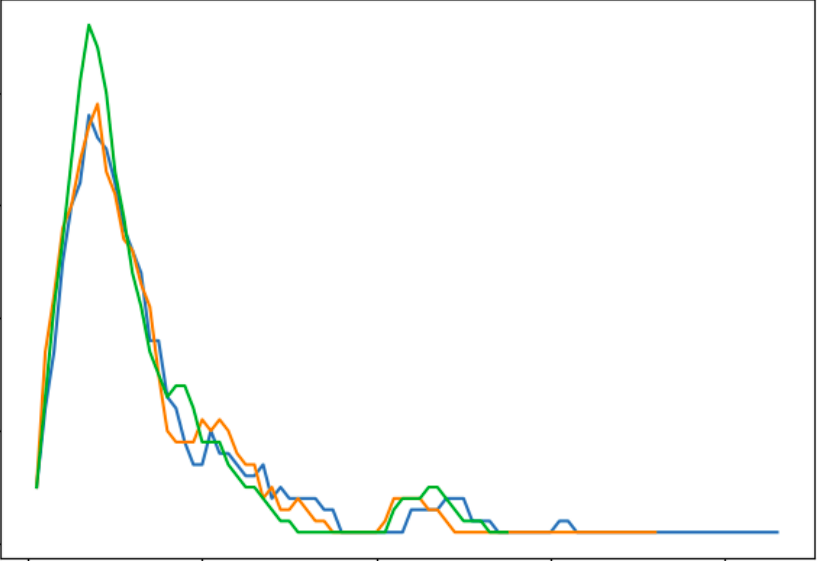};
\end{axis}
\end{tikzpicture}
\caption{SimPLoID output graphs showing the progress of the number of infected individuals over time from different settings (scatter plot vs line graph, 1 vs 5 simulations, short vs long period of resistance)}
\label{fig:graphs}
\end{figure}

The ProbLog code generated by the current implementation is still in parts more verbose than the code example given in Listing \ref{lst:ProbLog}, but it follows closely the structure illustrated there. The main deviation is that for performance reasons, our implementation explicitly grounds out the time variable in a preprocessing stage since the general grounding algorithm in ProbLog proved to be an efficiency bottleneck for larger simulations.  

\section{Discussion and outlook}\label{sec:Discussion}

Using probabilistic logic programming as a compilation target for
the DSL has several distinct advantages. 

Since the target language is itself a high-level probabilistic modelling language, the ease of configuration and individual extension that motivate the use of the DSL carry over to the SimPloID program itself. New language constructs in the DSL can be added in a modular fashion, with minimal interference in the existing translation. 

The compilation method also allows the generated ProbLog models to be inspected by programmers, knowledge engineers or experienced operators. As they are still human-readable and have a clearly-defined, declarative semantics, any discrepancy between the compiled model and its intended meaning can be pinpointed, allowing for the quick detection of bugs. These advantages are facilitated by the first-order relational nature of ProbLog, which keeps the compiled models very compact and allows for a clean separation of the disease model and the additional data, such as the supplied contact graph.

Indeed, the semantics of ProbLog clauses is ideally suited to network-based modelling through its intuitive implicit noisy-or combination function. In this way, if a single individual is exposed to two different possible sources of infection during a single time step, their probabilities of infection are compounded as independent events. Conversely, ProbLog also supports inhibition effects \cite{Meert2014}, which provide a concise and easily readable syntax for conditions which prevent an event, say an infection, from happening in a given timestep or globally.    

This is contrasted with the finite state machines used as a target formalism for the DSL in EMULSION, which are much more naturally suited to expressing compartmental models. 
When expressing network-based models, the EMULSION framework requires the user to encode explicit aggregation functions incorporating all possible effects, severely impacting the modularity of the specification and putting a significant burden on the modeller. 
Additionally, the lack of an overarching relational framework makes the large set of finite state machines dependent on the precise set of individuals, as well as being more difficult to inspect and reason about. 

Finally, compiling to a single ProbLog program allows for the future use of sophisticated static analysis tools such as probabilistic independence tests \cite{Rueckschloss2023} and asymptotic evaluation \cite{Quach2023}.

Among probabilistic logic programming languages, ProbLog is simple in its syntax and supported by two well-maintained implementations, cplint (in Prolog) and ProbLog 2 (in Python). Beyond simulation, they provide a variety of sophisticated learning and inference modes which could be added in future versions without altering the process generating the underlying ProbLog model. 
For instance, there are current plans to incorporate a learning mode which allows the estimation of disease
parameters from observational data, which would ordinarily require a complete rewrite of the software.

The main price to be paid for our approach is in the mediocre performance of the MCMC sampling process. 
This has two main causes. 
Firstly, the Python implementation of MCMC sampling in ProbLog 2 is optimised for ease of use rather than performance, as has been demonstrated empirically in the past \cite{Riguzzi2013}. 
Secondly, the code automatically generated by our translator is itself not ideal, since it does not cleanly separate deterministic and stochastic predicates. 

This is particularly problematic since in our experience, deterministic predicates tend to perform rather poorly under ProbLog 2 compared to state-of-the-art Prolog systems. 

Therefore, work is ongoing to enhance the performance by transitioning from ProbLog 2 to a variant of the cplint system, more specifically to a sampling algorithm adapted from MCINTYRE \cite{Riguzzi2013} under XSB Prolog. This is made possible by the recent release of the Janus bridge \cite{AndersenSwift2023}, which allows for a seamless integration of XSB Prolog from Python and thereby mitigates the downside of using a system from outside the host programming language of SimPloID.
The availability of a competitive tabled Prolog engine such as XSB would also obviate the need for manually grounding out the time variable in a preprocessing step, keeping the programs passed to the engine compact. 
On the other hand, porting our application also requires adding support for inhibition effects by reimplementing the syntactic transformation described by Meert and Vennekens \cite{Meert2014}. 

Additional directions for further work include support for multiple
simulations to be performed simultaneously using multiprocessing as well as integration of different levels of granularity. 
On an even finer scale, this extends to agent-based modelling, which has been prototyped as a separate application but is yet to be integrated coherently with the SimPloID framework presented here.

Seamless integration of the analysis tools for probabilistic independence and asymptotic behaviour is also intended in future releases.

\bibliographystyle{splncs04}
\newpage
\bibliography{Epidemiology_Almiftah}

\end{document}